\documentstyle[11pt]{article}
\begin{document}
 
\begin{center}
 
{\bf On the 3D Velocity Reconstruction of Clusters of
Galaxies}
\vspace{0.1in} 
 
V.G.~Gurzadyan$^1$ and S.~Rauzy$^2$
\vspace{0.1in}	
 
1. Department of Theoretical Physics, Yerevan Physics Institute, Yerevan
375036, Armenia;

2. Universit\'e de Provence and	Centre de Physique Th\'eorique -
C.N.R.S., Luminy Case 907, F-13288 Marseille
 Cedex 9, France.
\end{center}

\vspace{0.1in}

1. The reconstruction of the 3D-velocities of the clusters of galaxies
and their large-scale motions remains one of the central problems of
observational cosmology (see [\ref{FL},\ref{Rau95}] and references therein).
The reason is clear: these properties should
contain crucial information on the mechanisms of formation of the filaments 
and even on more early phases of evolution of the Universe, as well as on the
present values of the cosmological parameters.

Below we are formulating the problem of the 3D-velocity reconstruction of   
the cluster of galaxies based essentially on the measured redshift 
i.e. 1D-velocity distribution of the galaxies within the cluster. 
The procedure we propose includes the following steps:
a) the determination of the physical cluster;
b) the obtaining of the galaxy redshift
distribution of that cluster; c) the reconstruction of the mean 3D
velocity distribution of the cluster from
the redshift distribution of galaxies.

The first step can be performed by S-Tree
[\ref{Gur94a},\ref{Gur94b}]
 or similar technique
enabling one the separation of the physically interacting galaxies
from the 2D-image of the cluster area. 
We illustrate the proposed reconstruction procedure by 
numerical simulations, thus revealing the possibilities and the
limitations posed by the observational parameters of the clusters of galaxies. 

2. In 1935 Ambartsumian [\ref{Amb36}]
has solved the stellar dynamical problem of reconstruction 
of 2D and 3D velocity distributions based on the observed
line-of-sight velocity distributions of stars. The main assumption made
was the independence of the distribution  functions on
the spatial regions (directions).
Ambartsumian's formula relating the 3D velocity distribution
function $\phi(v_x,v_y,v_z)$ with the observed line-of-sight velocity
distribution $f(v_r,l,b)$ has the form:
\begin{equation}
\label{ambartsumian}
\begin{tabular}{l}
$\phi(v_x,v_y,v_z)= -\frac{1}{8\pi^2}\int dW\frac{1}{W}
\frac{d}{dW}$ \\
\\
$\int 
\frac{\cos b\, dl\, db}{\eta(l,b)}
f(v_x \cos l \cos b + v_y \sin l \cos b
+v_z \sin b+W,l,b)$
\end{tabular}
\end{equation}
where $W=v_r-x\cos\alpha-y\sin\alpha$ in some frame.
Computer experiments show that the direct application of the this
formula is hardly possible for that purpose: the given distribution for
the numerically simulated clusters does not coincide with the 
reconstructed one by means of that formula. The reason is clear: the
derivation of a smooth function based on  discrete
information on relatively small number of points ($10^2-10^3$) in a
nonlinear problem.
This fact is a consequence of  the principal difference
between the N-body problem in stellar dynamics and dynamics of clusters
of galaxies. 

However, we notice that the Ambartsumian's formula
has an interesting feature - the
radial velocity parameter $W$ is entering  into it in a similar 
way as the coordinates.   
Therefore an additional physically reasonable information  
on the features of the function $\phi(v_x,v_y,v_z)$
can make  the problem  correctly formulated and much
stable with respect to the initial noise. In the following,
we explore such an approach.

3. Our assumption is the representation of
the 3D velocity distribution in the following form:
\begin{equation}
\label{dpvisotropic}
\phi(v_x,v_y,v_z)~dv_x\,dv_y\,dv_z=\prod_{i=1}^3~
g_i(v_i;v_0^i,\sigma_v^i)
\,dv_i
\end{equation}
with $g_i(x;x_0,\sigma_x)$ being a smooth probability density
function centered on $x_0$ and of dispersion $\sigma_x$.
The  peculiar velocity field within the cluster  is thus
split into a mean $3$-dimensional velocity ${\bf v}_c=(v_0^1,v_0^2,v_0^3)$
plus random components of velocity dispersion $\sigma_v^i$. Herein,
we consider the isotropic case, choosing
$g_i\equiv g$ gaussian, i.e : the random component has a 3D Maxwellian
distribution of dispersion $\sigma_v$.

Note that the peculiar radial velocity $v_r={\bf {\hat r}.v}$
of the galaxies is not completely furnished
by the observed redshift $z= H_0\,r+{\bf {\hat r}.v}$, with $H_0$
the Hubble's constant.
Instead of usually made assumption that all galaxies lie at the same
distance we prefer a
more realistic one, permitting the cluster to have a spatial
line-of-sight extension $\sigma_c$, with
galaxies isotropically distributed around the center of the cluster.

Then, we rewrite the theoretical density
probability in terms of the observable variables $z$,
$l$ and $b$.
The successive integrations over the distance $r$ and over 2 components of
the 3-dimensional velocity field give the following observed probability
density :

$$dP_{\rm obs}=g\left(\right.z-H_0\,r_c;v_1 \cos l \cos b - v_2 \sin l \cos b 
- v_3 \sin b,$$ 
$$\left. \sqrt{\sigma_v^2+\sigma_c^2}
 \right)~\eta(l,b)~\cos b\,dl\,db\,dz$$
The estimates of the parameters $v_1$, $v_2$ and $v_3$ are obtained
by the maximized
likelihood function with respect to $v_1$, $v_2$, $v_3$, $H_0\,r_c$,
$\sigma_v$
and $\sigma_c$.

For $A=\cos l \cos b - \langle \cos l \cos b \rangle,
B=\sin l \cos b - \langle \sin l \cos b \rangle,
C=\sin b - \langle \sin b \rangle,
D=z - \langle z \rangle$,
where $\langle . \rangle$ denotes the average on the sample,
we obtain the following system of linear equations:

\begin{equation}
\label{system}
\left[ 
\begin{tabular}{ccc}
$\langle A^2 \rangle$ &
$\langle A.B \rangle$ &
$\langle A.C \rangle$\\
$\langle B.A \rangle$ &
$\langle B^2 \rangle$ &
$\langle B.C \rangle$\\
$\langle C.A \rangle$ &
$\langle C.B \rangle$ &
$\langle C^2 \rangle$
\end{tabular} 
\right]
\left[ 
\begin{tabular}{l}
$ v_1 $\\
$ v_2 $\\ 
$ v_3 $ 
\end{tabular} 
\right]~=~
\left[ 
\begin{tabular}{l}
$\langle A.D \rangle$ \\
$\langle B.D \rangle$\\
$\langle C.D \rangle$
\end{tabular} 
\right]
\end{equation}

One can see that the spatial line-of-sight expansion $\sigma_c$ is added
quadratically to the peculiar velocity dispersion $\sigma_v$,
enhancing thus the redshift scatter $\sigma_z$ (in km s$^{-1}$)
of the cluster. Moreover, because the distance $r_c$ of the cluster
normally is also an unknown
parameter, it is impossible to disentangle the estimates
of $H_0\,r_c$ and of the mean radial velocity $v_{rad}$
($v_{rad}= \langle {\bf {\hat r}} \rangle{\bf .v}$).
However it turns out that without the knowledge of $r_c$, one can
evaluate the mean tangential velocity ${\bf v}_{tan}$ of the cluster 
$$
{\bf v}_{tan}{\bf .}\langle {\bf {\hat r}} \rangle=0.
$$ 
We have estimated the accuracy of reconstruction of the mean tangential
velocity $\Delta v_{tan}$.
In figure 1, the results for the probability density
for velocity dispersion of  Maxwellian distribution of $\sigma_v=400$
km.s$^{-1}$ and spatial size  $\sigma_c=3$ Mpc are
represented.
It is interesting that this accuracy does not depend on the simulated
velocity ${\bf v}_c$ of the cluster.

\vspace{0.1in}

4. The proposed reconstruction procedure as revealed by the 
numerical experiments can  have a  range of
application for obtaining information on large-scale motions of
the filaments, if especially combined with the considerations on the possible
relation of the Hausdorff dimension of the
cluster and its dynamical properties [\ref{Gur91},\ref{Mon94}].

For a given velocity dispersion
the accuracy of the reconstructed transverse velocity
is well fitted by the following formula:
\begin{equation}
\label{relativistic}
\Delta v_{tan}=\frac{A(\sigma_v)}{\sqrt{N}}~ \frac{c\Psi(z)}{H_0}
\end{equation}
where for example 
$\Psi(z)=2/\Omega\,(\Omega z + (\Omega - 2)[\sqrt{1+\Omega z}-1])$
if $k=-1$ and tends to $z$ for small redshifts, 
$A(\sigma_v)=\sqrt{\sigma_c^2+\sigma_v^2}$, thus relating
the reconstructed $\Delta v_{tan}$ with the Hubble constant and other
 cosmological parameters.

V.G.Gurzadyan was supported by French-Armenian PICS.


\begin{thebibliography}{}

\bibitem{1}
\label{Amb36}
Ambartzumian V.A., M.N.R.A.S. {\bf 96}, 172 (1936).

\bibitem{2}
\label{FL}
Fisher K.B., Lahav O., Hoffman Y., Lynden-Bell D., Zaroubi S.,
M.N.R.A.S., {\bf 272}, 885 (1995).

\bibitem{2}
\label{Gur91}
Gurzadyan V.G., Kocharyan A.A., Europhys.Lett. {\bf 15}, 801 (1991).

\bibitem{3}
\label{Gur94a}
Gurzadyan V.G., Kocharyan A.A., '{\it 
Paradigms of the Large-Scale Universe}',
Gordon and Breach (1994).

\bibitem{4}
\label{Gur94b}
Gurzadyan V.G., Harutyunyan V.V., Kocharyan A.A., A\&A {\bf 281}, 964 (1994).

\bibitem{5}
\label{Mon94}
Monaco P., A\&A {\bf 287}, L13 (1994).

\bibitem{8}
\label{Rau95}
Rauzy S., Lachi\`eze-Rey M. and Henriksen R. N., Inverse Problems, (1995,
in press).

\end{thebibliography}
\end{document}